\documentclass[10.8pt,aps,prc,twocolumn,floatfix]{revtex4-1} 
\usepackage{graphicx}
\usepackage{sidecap}
\usepackage{array}
\usepackage{subfigure}
\usepackage{color}
\usepackage[columnwise]{lineno}
\usepackage{amsmath}
\usepackage{makecell}

\begin{document}
\title{Selecting specific initial configuration using spectator neutrons in U+U collisions}

 \author{Vipul Bairathi, Md. Rihan Haque,  Bedangadas Mohanty}
 \affiliation{National Institute of Science Education and Research,
   Jatni-752050, India} 

\begin{abstract}
\noindent We present a method to select events with specific initial 
configuration, namely body-tip, in heavy-ion collisions using deformed 
Uranium nuclei. We propose to use asymmetry in spectator neutron
numbers to filter out these body-tip events from the unbiased
configurations in U+U collisions. We have used a variable $S_{\eta}$ to
differentiate between the body-tip and unbiased configurations. 
We have calculated the 2$^{nd}$ order azimuthal anisotropy, namely
elliptic flow ($v_{2}$), for this body-tip configuration in the
framework of a transport model and found it to be consistently lower
compared to that in unbiased configurations as we expected. The purity
of selecting such events in a real experiment is also discussed.
\end{abstract}
\maketitle

\section{Introduction}
\noindent In a central heavy-ion collisions with spherical nuclei such as 
$Au$ or $Pb$, the initial overlap region is always circular. In the U$+$U
collisions, however, the initial overlap region can acquire different 
configurations owing to the deformed shape of the Uranium
nuclei~\cite{UU_paper}. In Ref.~\cite{UU_paper}, the elliptic flow
($v_{2}$) has been studied and found to be strongly correlated
with the different configurations of the initial overlap region. 
Moreover, in heavy-ion collisions, the energetic spectator protons 
can produce a strong magnetic field reaching 
$eB_{y} \sim$$m_{\pi}^{2}$~\cite{magfield}. Such a strong magnetic
field can give rise to chiral magnetic effect (CME) and chiral
separation effect (CSE)~\cite{magfield,CME}. However, to observe these
phenomena in real data, the azimuthal anisotropy ($v_{2}$) has to be
minimised as it acts as a background~\cite{STAR_CSE} to these processes.
Therefore, the events which have very high magnetic field and low
azimuthal anisotropy ($v_{2}$) are the perfect candidates for the
CME. In a central Au$+$Au or Pb$+$Pb collisions, the azimuthal 
anisotropy is very low, but the magnetic field is also low due to less
number of spectator protons. In a non-central Au$+$Au or Pb$+$Pb
collisions, although the magnetic field is comparatively high, the
azimuthal anisotropy also ($v_{2}$) starts increasing and therefore,
increasing the background to CME.  

\vspace{0.18cm}
\begin{figure}[!ht]
\centering
\includegraphics[scale=0.4]{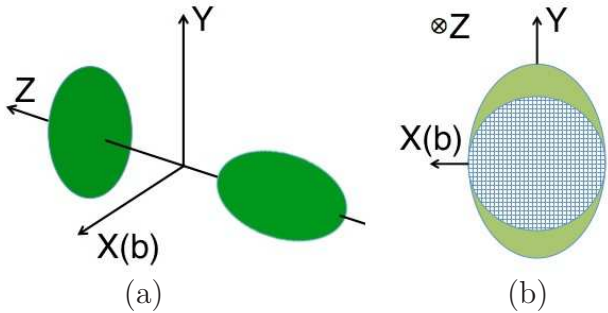}
 \put(-205,-5){\large{(a)}} 
 \put(-60,-5)  {\large{(b)}} 
\caption{\small{(a) (Color online) Body-tip configuration in the
    laboratory frame of reference. The impact parameter is along the
    X-axis. (b) The cross sectional view of the overlap region (shown
    by mesh) for a central ($b$ = 0) body-tip collision.} }
\label{fig:body_tip}
\end{figure}
\noindent Due to the deformed shape, the U$+$U collisions can
have unique orientations in which the magnetic field is very high  
in central collisions and as well as the azimuthal anisotropy is also very
low. Therefore, U$+$U collisions may provide an unique 
opportunity to study these exotic effects in relativistic heavy-ion 
collisions. However, it has not been experimentally possible so far to 
unambiguously select specific configurations in U+U collisions. 

In this paper, we propose a methodology to select body-tip 
configuration from unbiased events in U+U collisions. The body-tip 
configuration is pictorially shown in Figure~\ref{fig:body_tip}(a), 
where the impact parameter ($b$) is along X-axis and the beam 
direction is along Z-axis. In this configuration, the (right going)
uranium nuclei whose major axis is perpendicular to beam 
is called body and other one (left going) whose major axis is
along the beam is called tip~\cite{S_Chatt}. 
As seen in Figure~\ref{fig:body_tip}(b), the overlap region
in such a body-tip collisions is circular (shown by mesh). The nucleons
which lies in the overlap are called participants and those which lies
outside the overlap region and do not take part in the collisions 
are called spectators.  It is visible from Figure~\ref{fig:body_tip}(b) 
that one uranium nucleus get completely occluded into the other, 
leaving almost no spectators, where as the other one will always have 
some spectator from the non-overlapping regions. This gives rise 
to asymmetry in the spectator counts in the two opposite directions. 
We use this particular feature of this body-tip event configuration to 
separate it out from rest of all the other random configurations
possible in deformed uranium nuclei.

\begin{figure*}[t!]
\centering
\includegraphics[scale=0.32]{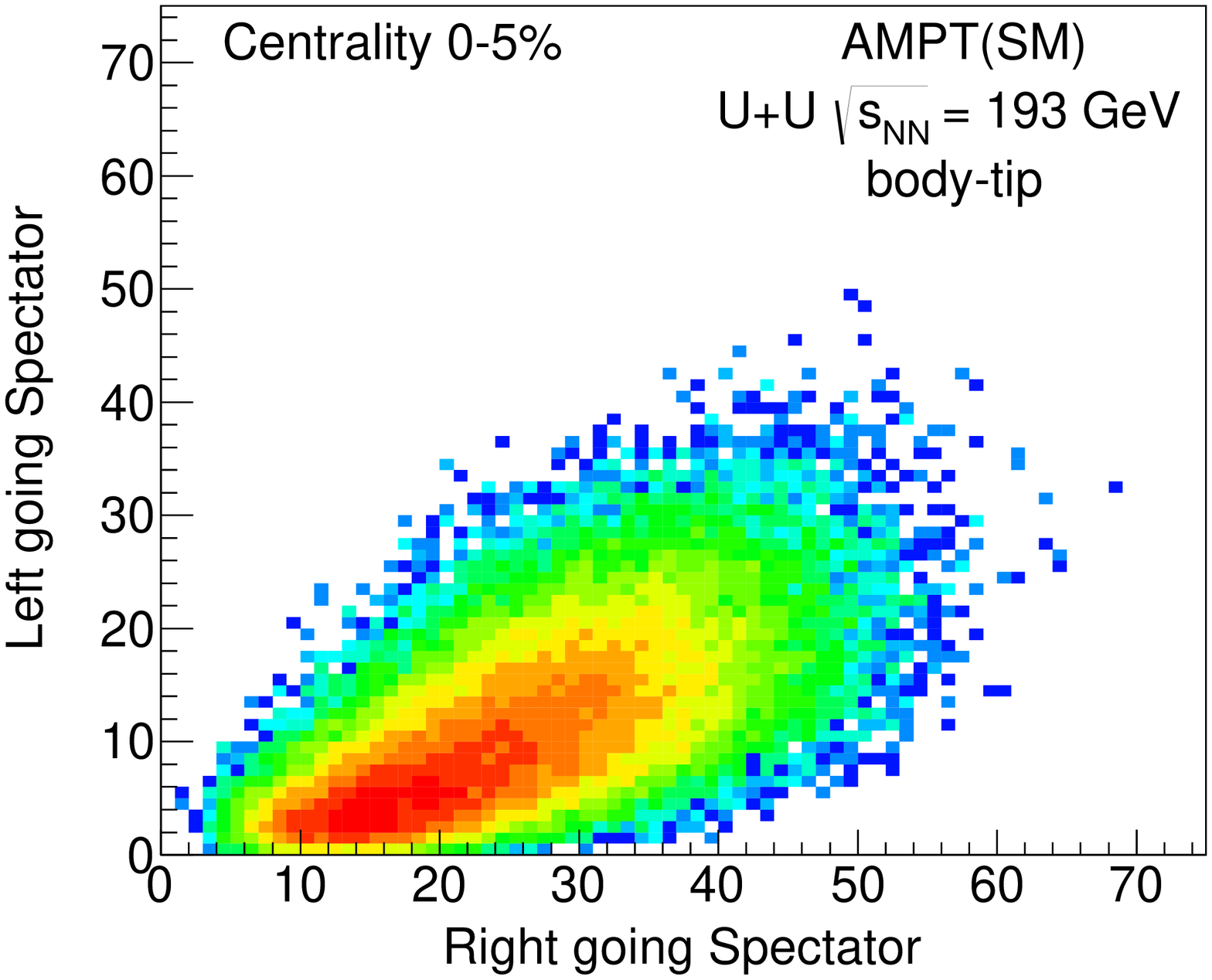}
 \put(-40,100){\large{(a)}} 
 \hspace{0.5cm}
\includegraphics[scale=0.32]{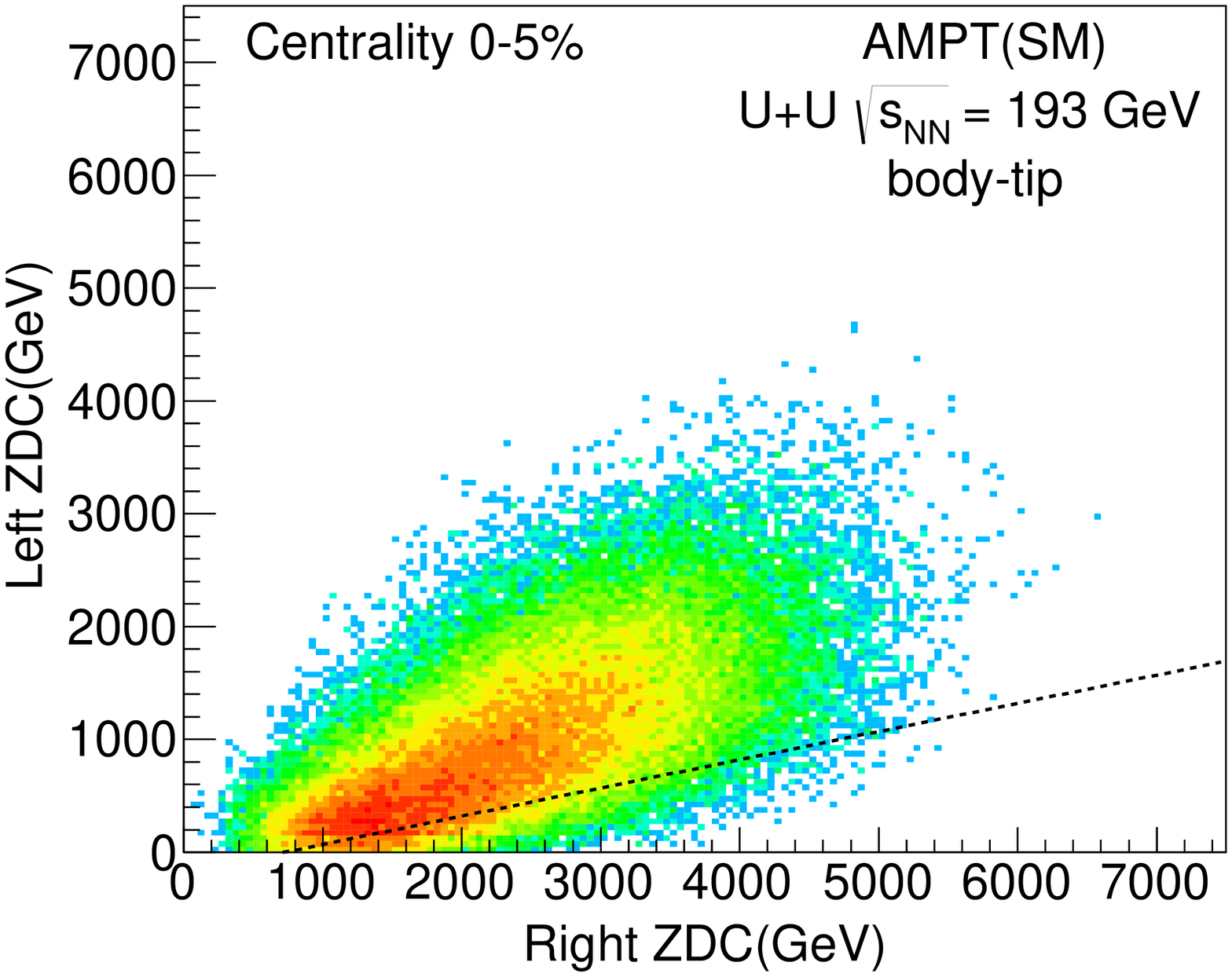}
 \put(-40,100){\large{(b)}} \\
 \vspace{0.28cm}
\includegraphics[scale=0.32]{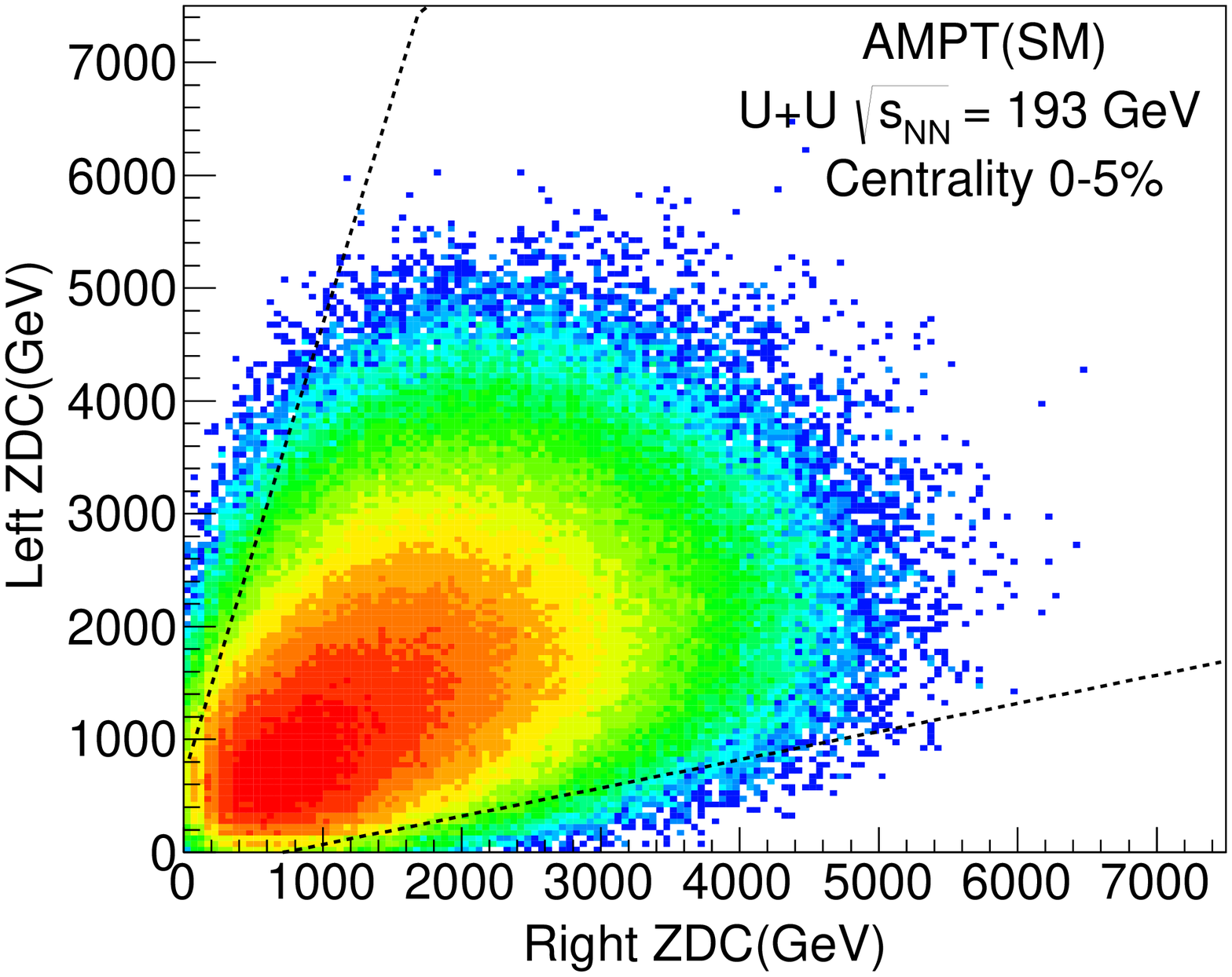}
\put(-40,100){\large{(c)}} 
 \hspace{0.5cm}
\includegraphics[scale=0.32]{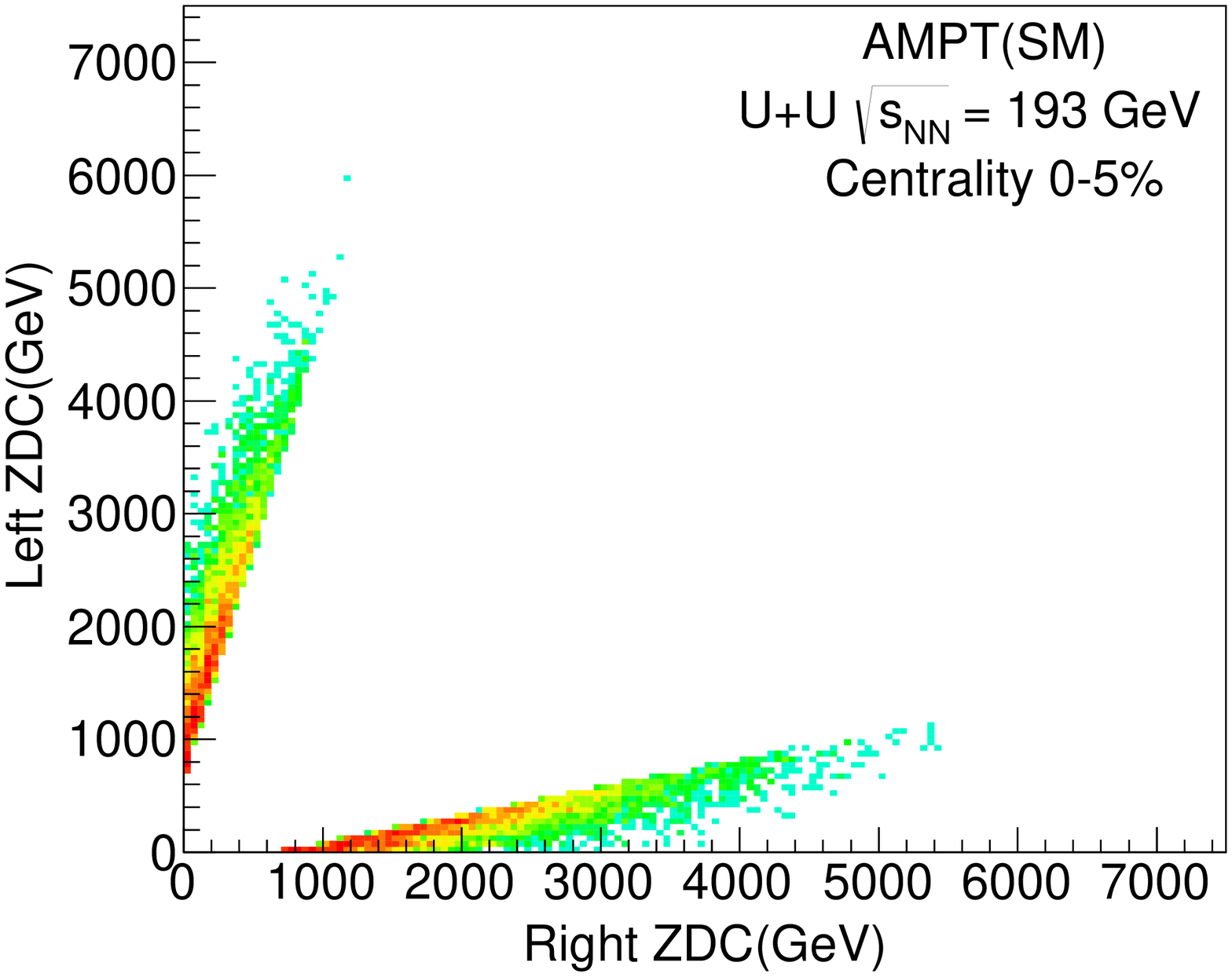}
\put(-40,100){\large{(d)}} 
\caption{\small{(a) (Color online) Spectator neutron distribution for
    0-5\% central events in body-tip configuration of U+U collisions
    at $\sqrt{s_{NN}}$ = 193 GeV. (b) Energy deposited  by spectator
    neutrons in ZDC for the same events of body-tip configuration as
    in (a). (c) Energy deposited by spectator neutrons in  ZDC for all
    configurations in U+U collisions. The dotted lines show the
    selection range for body-tip events. (d) The spectator neutron
    energy in ZDC after selecting events using the dotted lines in (c).}}
\label{fig:spec2d}
\end{figure*}

\vspace{0.5cm}

\section{The AMPT model}
\noindent We have used \textbf{A} \textbf{M}ulti \textbf{P}hase 
\textbf{T}ransport, namely the AMPT model, for our analysis.
 U+U collisions are implemented in the AMPT model by
deforming the Woods-Saxon profile~\cite{WoodSaxon} as,
\begin{equation}
 \rho\ = \frac{\rho_{0}}{1+exp([r-R^\prime]/a)}
 \label{eq:rho}
 \end{equation}
 \begin{equation}
  R^\prime\ = R[1+\beta_{2}Y^{0}_{2}(\theta)+\beta_{4}Y^{0}_{4}(\theta)]
 \label{eq:rho}
 \end{equation}
\noindent where $\rho_{0}$ is the normal nuclear density, $R$ is the 
radius of the nucleus, and $a$ denotes the surface diffuseness parameter.
$Y^{m}_{l}(\theta)$ denote spherical harmonics and $\theta$ is the
polar angle with respect to the symmetry axis of the nucleus. We have
used the values of $R$, $a$, $\beta_{2}$ and $\beta_{4}$
from~\cite{UU_paper}. The AMPT model, which is a hybrid transport
model, has four main stages: the initial conditions, partonic
interactions, the conversion from the partonic to the hadronic matter,
and finally hadronic interactions~\cite{ampt}. It uses the same
initial conditions from HIJING~\cite{hijing}. Scattering among partons
are modelled by Zhang’s parton cascade~\cite{ZPC}, which calculates
two-body parton scatterings using cross sections from pQCD with
screening masses. In the default AMPT model, partons are recombined 
with their parent strings and when they stop interacting, the
resulting strings fragment into hadrons according to the Lund string 
fragmentation model~\cite{lund}. However in the string melting (SM)
scenario, these strings are converted to soft partons and a quark 
coalescence model is used to combine parton into hadrons. The 
evolution dynamics of the hadronic matter is described by A 
Relativistic Transport (ART) model~\cite{ART}. We have used string 
melting (SM) mode of AMPT version 2.25t7, with parton-parton
cross section of 10 mb which will give rise to substantial amount of 
$v_{2}$. A total of 0.6 million 0-5\% central events were generated
using this model for the analysis.

\leftlinenumbers
\section{Results and discussion}
\noindent In the experiments it is possible to get the measure of
spectators using zero degree calorimeter (ZDC) detector 
which lies very close to the beam pipe in the forward (and backward)
direction. The ZDC detector gives an electrical signal which is
proportional to the number of spectator neutrons. Therefore, we
will use only the neutrons from the spectators for our study.
In Figure~\ref{fig:spec2d} we show the spectator neutron 
correlation for both body and tip oriented Uranium nuclei in body-tip
collisions. As seen from the figure, the spectator neutron counts are 
not symmetric for body-tip collisions. In a real experiment, these 
spectator neutrons are detected by Zero Degree Calorimeters (ZDC). 
Therefore, it is worthwhile to convert this spectator neutrons into 
practically measurable ZDC signal. We use the ZDC response of STAR 
experiment from Ref.~\cite{ZDC}. From Fig.6 of Ref.~\cite{ZDC} we
find that the resolution of single neutron ZDC response is 18\%. 
Therefore, we smear the energy deposited by each individual spectator
neutron by a Gaussian distribution with the width of 18\% of the mean
value to imitate the response in the ZDC detector. The mean value of
energy deposited by a single neutron for this study is 96.5 GeV. The
energy deposited by the spectator neutrons in the ZDC, event by event, 
is shown in Figure~\ref{fig:spec2d}(b). We can select the
body-tip events from all the other configurations using this correlation
shown in Figure~\ref{fig:spec2d}(b). One such selection procedure is
shown by the dotted line, with slope = 0.25 and intercept = -180. We
select all the events which lies below this line, therefore selecting the 
events with asymmetric spectator neutrons. The ZDC response for all
possible configurations in U+U collisions is shown in
Figure~\ref{fig:spec2d}(c). Since both left going and right going
nuclei can be in either body or in tip orientation, therefore,
we select these events along both (left and right) axis. Two dotted 
lines show the selection ranges for the possible body-tip events.  
Figure~\ref{fig:spec2d}(d) shows the ZDC response of the selected
events. 

\begin{figure}[h]
\centering
\includegraphics[scale=0.40]{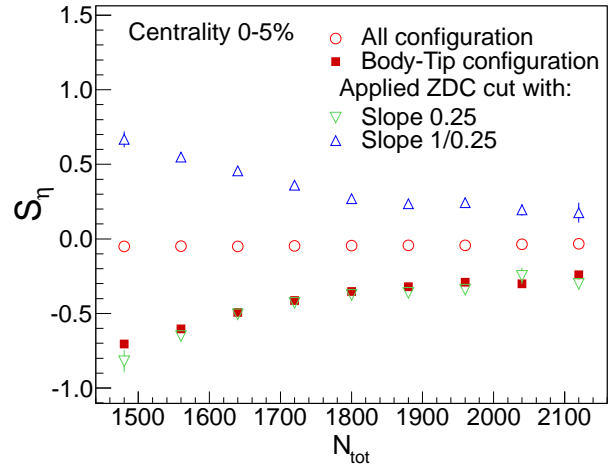}
\caption{\small{(Color online) $S_{\eta}$ (see
    Eqn.~\ref{eq:S_eta}) for minimum bias, selected body-tip events
   from minimum bias, and pure body-tip events. } }
\label{fig:seta}
\end{figure}

  One way to differentiate between the minimum bias (all possible
configurations) and the body-tip configurations is to look at the
variable $S_{\eta}$ which is defined as, 
 \begin{equation}
 S_{\eta}\ =  \frac{ \sum \eta (dN/d\eta)}{N_{tot}}
 \label{eq:S_eta}
 \end{equation}
where, $N_{tot}$ is the number of particles within pseudorapidity
range, -1.0 $< \eta <$ 1.0 and the summation is over all particles in
the event. 
Figure~\ref{fig:seta} shows the variable $S_{\eta}$ as a function of
$N_{tot}$ for minimum bias, events selected with cuts on the ZDC
signal from minimum
bias configuration, and pure body-tip events. The $S_{\eta}$ for minmum bias configurations
lies close to zero, suggesting symmetry in particle production. The
particle production in body-tip events are asymmetric in $\eta$ as
shown by solid squares in Fig.~\ref{fig:seta}. Whereas, when selecting
body-tip events from minimum bias configurations, both the projectile
and the target can either be in body or in tip configuration. Therefore 
we have two set of $S_{\eta}$ (when projectile is body, target is tip
and vice-versa) for the selected events. The difference observed in
$S_{\eta}$ of selected events and minimum bias configurations enhances
the possibility of our method to select the body-tip events in real
experiments.   

\begin{figure*}[t!]
\centering
\includegraphics[scale=0.35]{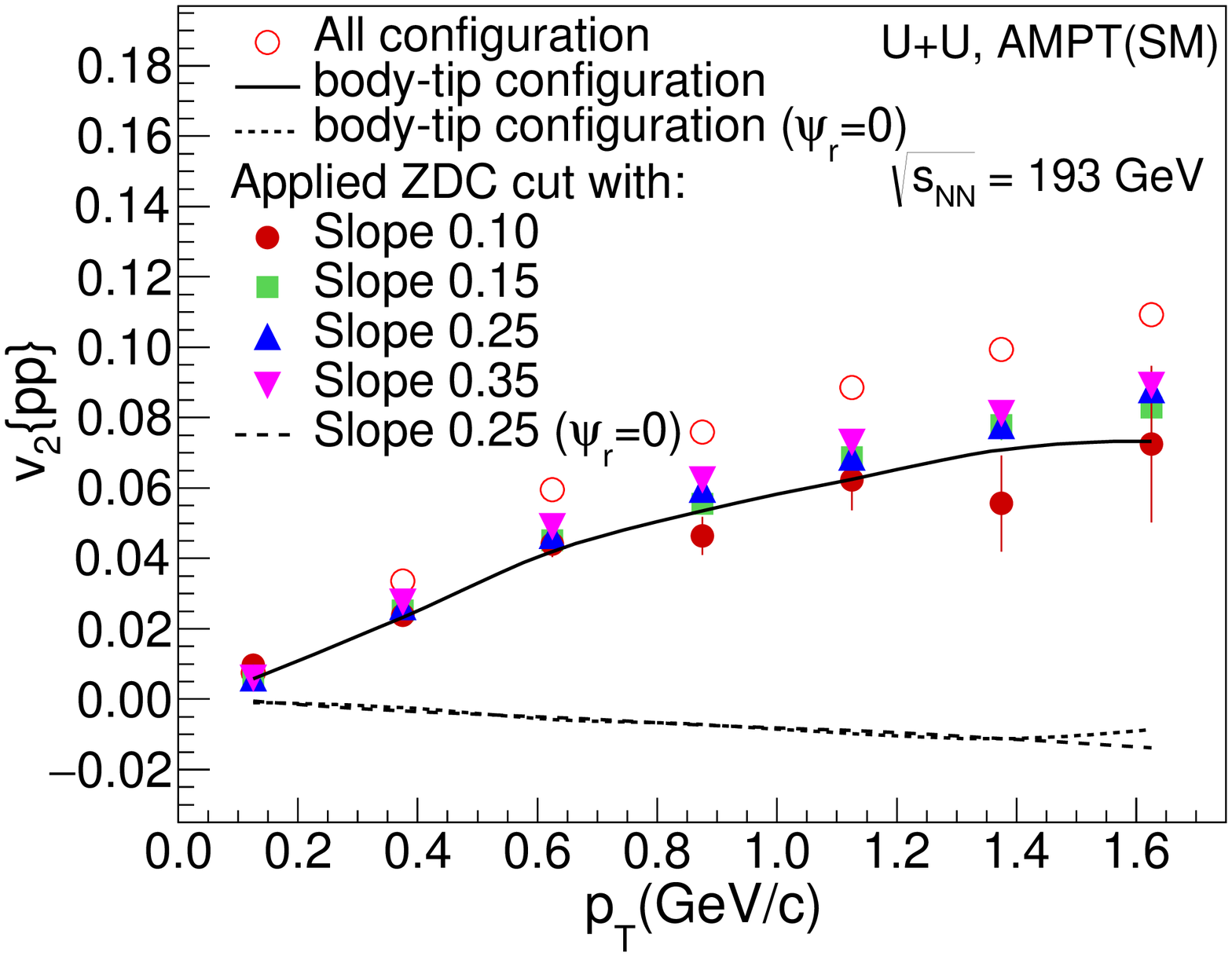}
 \put(-70,115){\large{(a)}} 
 \hspace{0.4cm}
\includegraphics[scale=0.35]{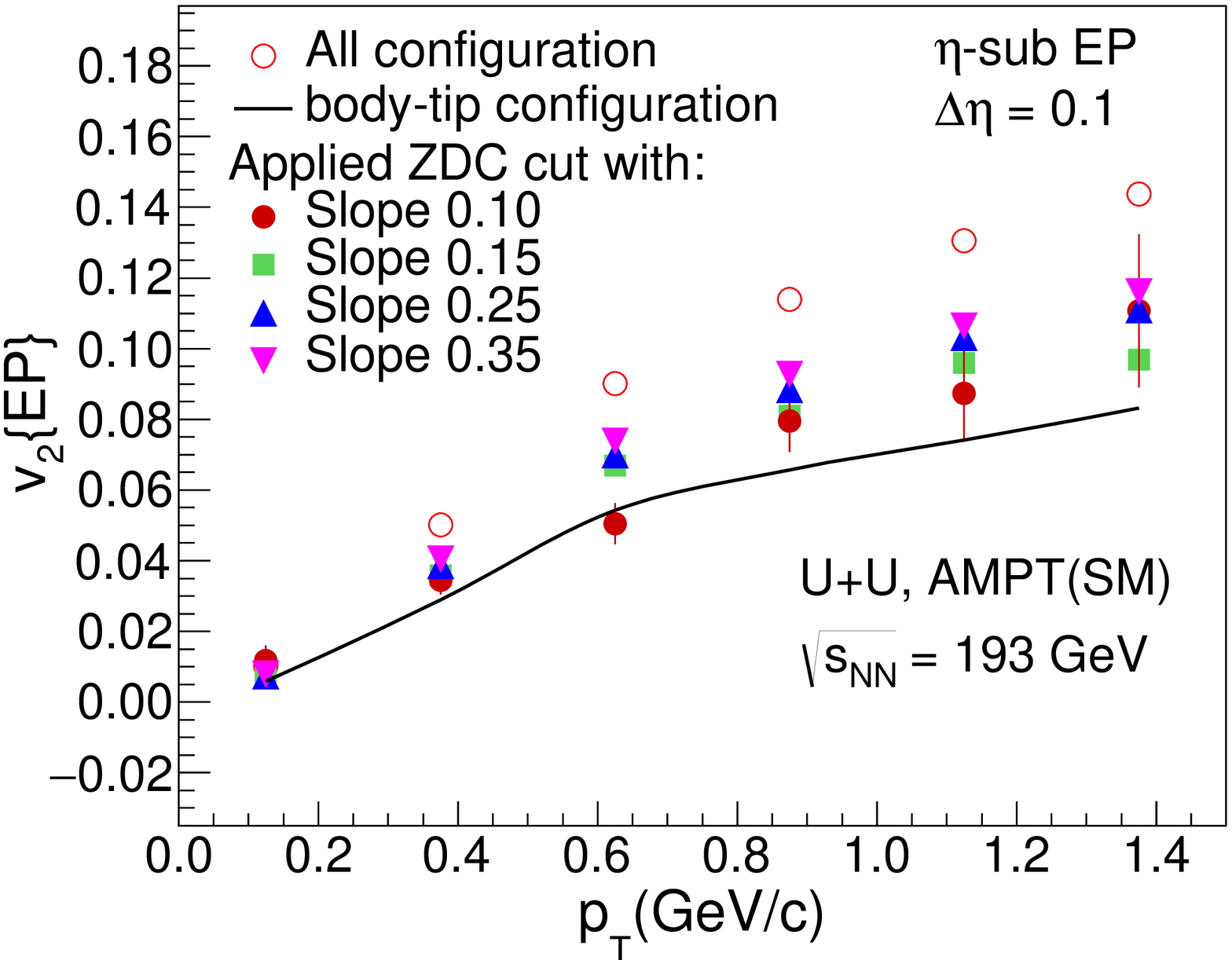}
 \put(-70,120){\large{(b)}} \\
 \vspace{0.42cm}
\includegraphics[scale=0.35]{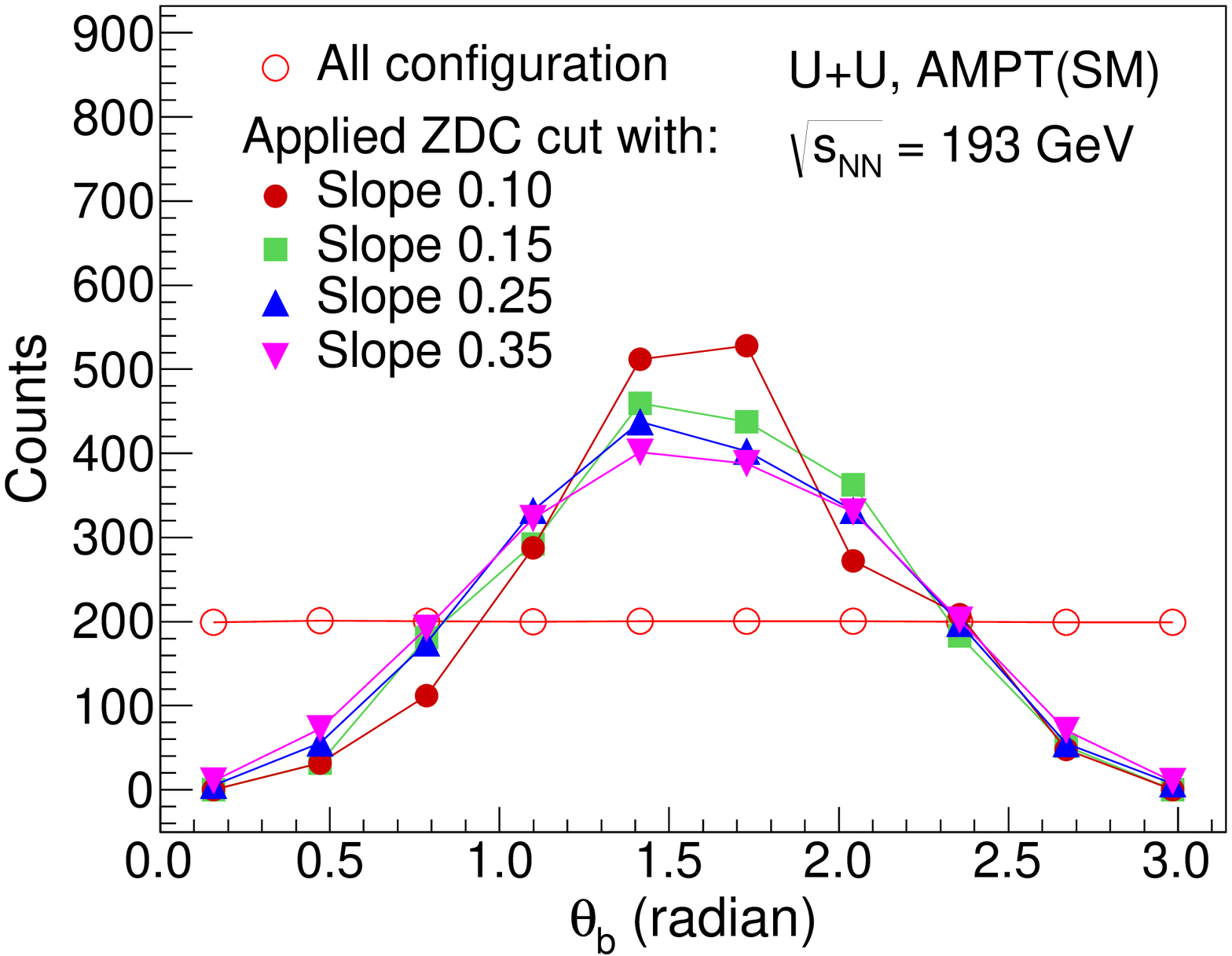}
 \put(-70,115){\large{(c)}} 
 \hspace{0.4cm}
\includegraphics[scale=0.35]{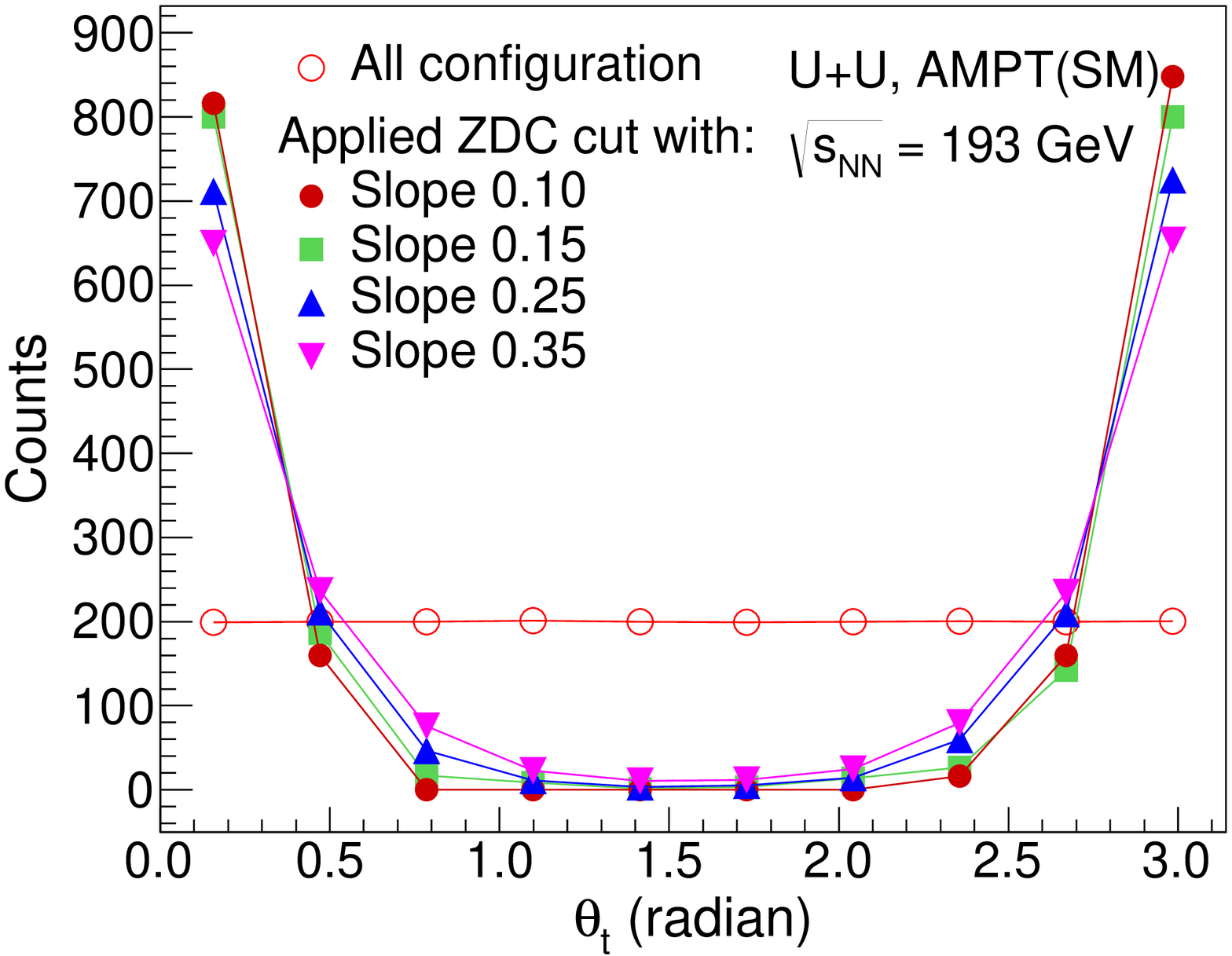}
 \put(-70,115){\large{(d)}} 
\caption{\small{(Color online) (a) $v_{2}\{pp\}$ for 0-5\% central U+U collisions at 
 $\sqrt{s_{\textrm {NN} }} =$ 193 GeV without ZDC cut (open markers) 
  and with ZDC cut (solid markers). (b) $v_{2}\{\textrm {EP}\}$ for
  the same events without ZDC cut (open markers) and with ZDC cut 
  (solid markers). Solid line in (a) and (b) corresponds to $v_{2pp}$
  and $v_{EP}$ for pure body-tip events. (c) Distribution of the angle
  of the major axis with the beam axis for body $(\theta_{b})$ and for
  (d) tip $(\theta_{t})$ oriented Uranium nuclei for different slope
  parameters. For pure body-tip events,
  $(\pi/2-0.005)<\theta_{b}<(\pi/2+0.005)$ and $0<\theta_{t}<0.05$
  (radian).}}
\label{fig:v2_all}
\end{figure*}

   Now that we think we have potentially selected body-tip events
from all configurations, we are going to look at $v_{2}$ of selected
events. As the overlap region in a central body-tip collision is
circular, therefore we expect that magnitude of the $v_{2}$ for
selected events should be less compared to a set of events with an
specific selection of configuration in U+U
collisions. Figure~\ref{fig:v2_all} shows the $v_{2}$ of charged
particles in mid-rapidity (-1.0$ <\eta< $1.0), measured with respect
 to participant plane $(\Psi_{pp})$ for 0-5\% central events
in U+U collisions at $\sqrt{s_{\textrm {NN} }} =$ 
 193 GeV. The detailed method of calculating $\Psi_{pp}$
using the position co-ordinates of participants (nucleons which take
part in the collisions) is given in Ref.~\cite{psi_pp}. From
Figure~\ref{fig:v2_all}(a) we see that the $v_{2}$ in body-tip events
is lower compared to minium bias configurations in U+U collisions. Now
the selected body-tip events might also contain some small amount of
other possible configurations of U+U collisions. Therefore, we have
varied the slope of the dotted line shown in Figure~\ref{fig:spec2d}(c) 
to make the selection more strict or relaxed. The corresponding change
in the $v_{2}$ is also shown in the Figure~\ref{fig:v2_all}(a). The
magnitude of charged particle $v_{2}(p_{T})$ in the selected body-tip
events 

\vspace{0.0cm}

\noindent are systematically 25\%
lower than that in all configurations. The $v_{2}(p_{T})$ for purely
body-tip events (i.e., orientation of Uranium nuclei are fixed
according to body and tip orientation) in AMPT is also shown as solid
curve in Figure~\ref{fig:v2_all}(a). The dotted and dashed lines in
Figure~\ref{fig:v2_all}(a) corresponds to the $v_{2}$ measured with
respect to reaction plane angle ($\Psi_{\textrm r}$) in body-tip and
selected body-tip events from all configurations
respectively. The difference in the $v_{2}\{\Psi_{\textrm r}\}$ and
$v_{2}\{\textrm {pp}\}$ shown in Figure~\ref{fig:v2_all}(a) arises due
the fluctuations in the initial participant distribution. The reaction
plane angle $\Psi_{\textrm r}$ is defined as the angle between the
impact parameter and the X-axis 
and is a known quantity in the AMPT model. In experiments, however,
the position of the participant nucleons are unknown. In this scenario,
the event plane is calculated using the anisotropic distribution of
the produced particles~\cite{art1}. We followed $\eta$ sub-event plane
method to calculate $v_{2}$ of charged particles. In this method, each
event is divided into two uncorrelated sub-events in two different
$\eta$ windows. Then 2$^{nd}$ order event plane ($\Psi_{2}$) is
calculated separately in both of these sub-events. Each particle is
then correlated with the event plane of opposite $\eta$ so as to
remove the self-correlation effect~\cite{art1}. The $v_{2}$ result
obtained in this method is then corrected for the $\eta$ sub-event plane
resolution~\cite{art2}. We have followed event-by-event resolution
correction~\cite{hiroshi_alex} for our analysis. The details of the
procedure of event plane calculation and resolution correction can be
found in~\cite{BES_v2}. The $v_{2}\{\textrm {EP}\}$ results for 0-5\%
central U+U collisions, calculated with event plane method, are shown in
Figure~\ref{fig:v2_all}(b). The $v_{2}\{\textrm {EP}\}$ 
for all configuration is shown by open markers in
Figure~\ref{fig:v2_all}(b). The $v_{2}\{\textrm {EP}\}$ for different 
ZDC cuts are shown by solid markers. As seen in
Figure~\ref{fig:v2_all}(b), $v_{2}\{\textrm {EP}\}$ is also systematically
25\%  lower for selected body-tip events compared 

\noindent 
to $v_{2}\{\textrm {EP}\}$ of all configurations. The $v_{2}\{\textrm
{EP}\}$ for purely body-tip events is shown by solid curve in
Figure~\ref{fig:v2_all}(b). As we see in Figure~\ref{fig:v2_all}(b),
with decreasing slope parameter (i.e., selecting events with higher
spectator neutron asymmetry),  $v_{2}\{\textrm{EP}\}$ of selected
events systematically decreases compared to that of all
configurations. Therefore, using lower slope value (higher spectator
asymmetry), we can enhance the selection of body-tip events from all
configurations in U+U collisions. Figure~\ref{fig:v2_all}(c) and Figure~\ref{fig:v2_all}(d)
shows the distribution of the angle of the major axis of the Uranium 
with the beam axis for body ($\theta_{b}$) and tip ($\theta_{t}$)
orientation for slope = 0.25. $\theta_{b}$ and $\theta_{t}$ for two
different slope values and for all configuration in U+U
collisions are also shown for comparison. 
Although lower slope values enhance the selection of body-tip events
from all configuration,  it also results in reduced event statistics. 
The selected events may possibly contain some other configurations.
Therefore, we have calculated the purity of the selected events. By
purity we infer that how much of the selected events can be specified
as body-tip type events.  Therefore, purity(\%) = (number of selected
events with $\theta_{t}$,\ $\theta_{p}$ within the limits of
relaxation $\theta_{R}$) /  (total number of events satisfying the ZDC
cut)$\times$100. Here we call an selected event to be body-tip if
-$\theta_{R}<\theta_{t}<\theta_{R}$ and $(\pi/2-\theta_{R})<\theta_{b}<(\theta_{R}+\pi/2)$.

\begin{table}[!h]
\caption{Purity (in \%) of selected events for different slope parameter and
  angular relaxation (see text for details).}
\vspace{0.2cm}
\begin{tabular}{|l|c|l|c|l|c|l|} 
\hline
\diaghead{\theadfont Keeping  this  space  for c }%
 {slope \\ parameter}{Angular \\ relaxation \\ $(\theta_{R})$} &
 \thead{$\pm$10$^{\circ}$} & \thead{$\pm$20$^{\circ}$} & \thead{$\pm$30$^{\circ}$}\  \\  [0.2ex]  
\hline
\ \ \  0.10   & \ 34\%  \  & \ 62\% \ & \ 74\%\ \ \\  [0.2ex] 
\hline
\ \ \  0.15   & \ 28\%  \  & \ 52\% \ & \ 70\%\ \ \\  [0.2ex] 
\hline
\ \ \  0.25   & \ 26\%  \  & \ 48\% \ & \ 67\%\ \ \\   [0.2ex] 
\hline
\ \ \  0.35   & \ 24\%  \  & \ 45\% \ & \ 63\%\ \ \\   [0.2ex] 
\hline 
\end{tabular}
\label{table:purity}
\end{table}

\noindent Since the angular distribution of both of the selected
Uranium nuclei has some finite width, therefore, we assume that out of
the selected events, the number of events for which $\theta_{b}$ and
$\theta_{t}$ lies within $\pm\theta_{R}$ degree of the corresponding
default values in body-tip events ($\theta_{b}$ = $\pi/2$ and
$\theta_{t}$ = 0) is the pure body-tip sample. It is worth mentioning
that the purity of the selected events depends on the relaxation on
the angular width ($\theta_{R}$) we set to classify the selected
event as body-tip. Hence we calculated the purity of the selected
event sample for different angular relaxation and the results are
listed in Table~\ref{table:purity}. As seen from
Table~\ref{table:purity}, the purity of selected events increases as
we decrease the value of slope parameter or if we increase the angular
relaxation. As can be found in Table~\ref{table:purity}, more than
70\% purity can be achieved using low slope value (i.e., higher order
of spectator neutron asymmetry).

\section{Summary} 
\noindent We present an experimental procedure to select the body-tip
configuration among all possible configuration in 0-5\% central U+U 
collisions at $\sqrt{s_{NN}}$ = 193 GeV. We found that the spectator
neutron energy deposited in the Zero Degree Calorimeter (ZDC) is a
useful tool to select body-tip oriented events in central U+U
collisions. We are able to select body-tip configuration with
conditions applied on spectator neutron asymmetry simulated with the ZDC. 
We have used a new variable $S_{\eta}$ to differentiate
between the body-tip and the minimum bias configurations. 
Elliptic flow ($v_{2}$) is calculated for the selected events with
respect to both participant plane angle ($\Psi_{pp}$) and event plane
angle ($\Psi_{2}$). As expected, $v_{2}$ of selected events is found to be systematically
lower compared to that in all configurations in U+U collisions. The ZDC
selection cut (slope) was varied and it was found that selecting
events with higher spectator neutron asymmetry results in lower
$v_{2}$ values which tends to match with $v_{2}$ of pure body-tip
events. Finally we calculated the purity of the selected events from
all configurations in the U+U collisions. We observed that purity
increases for decreasing slope parameter. In other words, if we apply
cut selecting higher order of left-right spectator neutron asymmetry,
then the purity of the selected events increases.

 \section*{Acknowledgement}
 \noindent We acknowledge our fruitful discussions with
 Dr. S. Chatterjee. This work is supported by the DAE-BRNS project
 Grant  No. 2010/21/15-BRNS/2026.

\end{document}